\documentclass[12pt, amsfonts]{article}
\begin{document}
\def\p {{\partial}}
\def\n {{\nu}}
\def\m {{\mu}}
\def\a {{\alpha}}
\def\bt {{\beta}}
\def\f {{\phi}}
\def\th {{\theta}}
\def\g {{\gamma}}
\def\eps {{\epsilon}}
\def\e {{\psi}}
\def\k {{\chi}}
\def\la {{\lambda}}
\def\na {{\nabla}}
\def\bn {\begin{eqnarray}}
\def\en {\end{eqnarray}}
\title{The equivalence between the Hamiltonian and Lagrangian formulations for
the parametrization invariant theories\footnote{e-mail:
$sami_{-}muslih$@hotmail.com}} \maketitle
\begin{center}
\author{S. I. MUSLIH\\\it{Department of Physics Al-Azhar University
Gaza, Palestine}}
\end{center}
\hskip 5 cm

$\mathbf{Summary}$.- The link between the treatment of singular
Lagraingians as field systems and the the canonical Hamiltonian
approach is studied. It is shown that the singular Lagrangians as
field systems are always in exact agreement with the canonical
approach for the parametrization invariant theories.


 Ef- Lagrangian and Hamiltonian approach.

\newpage

\section{Introduction}

In previous papers [1-4] the Hamilton-Jacobi formulation of
constrained systems has been studied. This formulation leads us
to obtain the set of Hamilton-Jacobi partial differential
equations [HJPDE] as follows:

\bn &&H^{'}_{\a}(t_{\bt}, q_a, \frac{\p S}{\p q_a},\frac{\p S}{\p
t_{\a}}) =0,\nonumber\\&&\a, \bt=0,n-r+1,...,n, a=1,...,n-r,\en
where
\begin{equation}
H^{'}_{\a}=H_{\a}(t_{\bt}, q_a, p_a) + p_{\a},
\end{equation}
and $H_{0}$ is defined as
\bn
 &&H_{0}= p_{a}w_{a}+ p_{\m} \dot{q_{\m}}|_{p_{\n}=-H_{\n}}-
L(t, q_i, \dot{q_{\n}},
\dot{q_{a}}=w_a),\nonumber\\&&\m,~\n=n-r+1,...,n. \en

The equations of motion are obtained as total differential
equations in many variables as follows:

\bn
 &&dq_a=\frac{\p H^{'}_{\a}}{\p p_a}dt_{\a},\;
 dp_a= -\frac{\p H^{'}_{\a}}{\p q_a}dt_{\a},\;
dp_{\bt}= -\frac{\p H^{'}_{\a}}{\p t_{\bt}}dt_{\a}.\\
&& dz=(-H_{\a}+ p_a \frac{\p
H^{'}_{\a}}{\p p_a})dt_{\a};\\
&&\a, \bt=0,n-r+1,...,n, a=1,...,n-r\nonumber \en where
$z=S(t_{\a};q_a)$. The set of equations (4,5) is integrable [3,4]
if

\bn &&dH^{'}_{0}=0,\\
&&dH^{'}_{\m}=0,\;\; \m=n-r+1,...,n. \en If condition (6,7) are
not satisfied identically, one considers them as new constraints
and again testes the consistency conditions. Hence, the canonical
formulation leads to obtain the set of canonical phase space
coordinates $q_a$ and $p_a$ as functions of $t_{\a}$, besides the
canonical action integral is obtained in terms of the canonical
coordinates.The Hamiltonians $H^{'}_{\a}$ are considered as the
infinitesimal generators of canonical transformations given by
parameters $t_{\a}$ respectively.

In ref. [5] the singular Lagrangians are treated as field systems.
The Euler-Lagrange equations of singular systems are proposed in
the form
\begin{equation}
\frac{\p }{\p t_{\a}}[\frac{\p L'}{\p (\p_{\a} q_{a})}]- \frac{\p
L'}{\p q_{a}}=0,\;\;\; \p_{\a}q_{a} = \frac{\p q_{a}}{\p t_{\a}},
\end{equation}
with constraints \bn&& dG_{0}= -\frac{\p L'}{\p t}dt,\\
&&dG_{\m}= -\frac{\p L'}{\p q_{\m}}dt, \en where \bn&& L'(t_{\a},
\p_{\a}q_{a}, \dot{q_{\m}}, q_{a})= L(q_{a}, q_{\a}, \dot{q_{a}}=
(\p_{\a} q_{a}){\dot{t_{\a}}}),\;\;\;\dot{q_{\m}}=
\frac{dq_{\m}}{dt},\\
&&G_{\a}= H_{\a}(q_{a}, t_{\bt}, p_{a}= \frac{\p L}{\p
{\dot{q_{a}}}}). \en

In order to have a consistant theory, one should consider the
variations of the constraints (9), (10).

In this paper we would like to study the link between the
treatment of singular Lagrangians as field systems and the
canonical formalism for the parametrization invariant theories.

\section{ Prametrization invariant theories as singular systems}

In ref. [3] the canonical method treatment of the
parametrization-invariant theories is studied and will be briefly
reviewed here.

 Let us consider a system with th action integral as
\begin{equation}
S(q_{i}) =\int dt {\cal L}(q_{i}, \dot{q_{i}},
t),\;\;\;\;i=1,...,n,
\end{equation}
where $\cal L$ is a regular Lagrangian with Hessian $n$.
Parameterize the time $t\rightarrow\tau(t)$, with $\dot{\tau}
=\frac{d \tau}{dt}>0$. The velocities $\dot{q_{i}}$ may be
expressed as
\begin{equation}
\dot{q_{i}}= q_{i}^{'}\dot{\tau},
\end{equation}
where $ q_{i}^{'}$ are defined as
\begin{equation}
 q_{i}^{'}= \frac{dq_{i}}{d\tau}.
\end{equation}
Denote $t= q_{0}$ and $ q_{\m}=(q_{0}, q_{i}),\;\; \m=0,
1,...,n,$ then the action integral (13) may be written as
\begin{equation}
S(q_{\m}) =\int d\tau  \dot{t}{\cal L}(q_{\m}, \frac{
q_{i}^{'}}{\dot{t}}),
\end{equation}
which is parameterization invariant since $L$ is homogeneous of
first degree in the velocities $ q_{\m}^{'}$ with $L$ given as
\begin{equation}
L(q_{\m}, \dot{q_{\m}}) = \dot{t}{\cal L}(q_{\m}, \frac{
q_{i}^{'}}{\dot{t}}).
\end{equation}
The Lagrangian $L$ is now singular since its Hessian is $n$.

The canonical method [1-4] leads us to obtain the set of
Hamilton-Jacobi partial differential equations as follows: \bn
{H'}_{0}=&& p_{\tau} -L(q_{0}, q_{i}, \dot{q_{0}}, {\dot{q_{i}}}=
w_{i}) + p_{i}^{\tau}q_{i}^{'} +\nonumber\\&&
p_{t}\dot{q_{0}}\mid_{p_{t}= -H_{t}}=0,
\;\;\;\;\;\;\;\;\;\;\;\;\;\;\;\;\;\;\;\;\;\;\;\;\;
\;\;\;\;\;\;\;p_{\tau}= \frac{\p S}{\p {\tau}},\\
{H'}_{t} =&& p_{t} +
H_{t}=0,\;\;\;\;\;\;\;\;\;\;\;\;\;\;\;\;\;\;\;\;\;\;\;\;\;\;\;\;
\;\;\;\;\;\;\;\;p_{t}=\frac{\p S}{\p t}, \en where $H_{t}$ is
defined as
\begin{equation}
H_{t}= -{\cal L}(q_{i}, w_{i}) + p_{i}^{\tau} w_{i}.
\end{equation}
Here, $p_{i}^{\tau}$ and $p_{t}$ are the generalized momenta
conjugated to the generalized coordinates $q_{i}$ and $t$
respectively.

The equations of motion are obtained as total differential
equations in many variables as follows: \bn&&dq^{i}= \frac{\p
{H'}_{0}}{\p p_{i}}d\tau + \frac{\p {H'}_{t}}{\p p_{i}}dq^{0}=
\frac{\p {H'}_{t}}{\p p_{i}}dq^{0},\\
&&dp^{i}=- \frac{\p {H'}_{0}}{\p q_{i}}d\tau + \frac{\p
{H'}_{t}}{\p q_{i}}dq^{0}= -
\frac{\p {H'}_{t}}{\p q_{i}}dq^{0},\\
&&dp_{t}=- \frac{\p {H'}_{0}}{\p q_{0}}d\tau + \frac{\p
{H'}_{t}}{\p q_{0}}dq^{0}= 0. \en Since
\begin{equation}
d{H'}_{t} = dp_{t} + H_{t},
\end{equation}
vanishes identically, this system is integrable and the canonical
phase space coordinates  $q_{i}$ and $p_{i}$ are obtained in
terms of the time $(q_{0}=t)$.

Now, let us look at the Lagrangian (17) as a field system. Since
the rank of the Hessian martix is $n$, this Lagrangian  can be be
treated as a field system in the form
\begin{equation}
q_{i}= q_{i}(\tau, t),
\end{equation}
thus, the expression
\begin{equation}
q_{i}^{'} = \frac{\p q_{i}}{\p \tau} + \frac{\p q_{i}}{\p t}{\dot
t},
\end{equation}
can be replaced in eqn. (17) to obtain the modified Lagrangian
$L'$:
\begin{equation}
L'= \dot{t}{\cal L}(q_{\m}, \frac{1}{\dot t}(\frac{\p q_{i}}{\p
\tau} + \frac{\p q_{i}}{\p t}{\dot t})).
\end{equation}
Making use of eqn (8), we have
\begin{equation}
\frac{\p L'}{\p q_{i}} - \frac{\p}{\p t}(\frac{\p L'}{\p
(\frac{\p q_{i}}{\p t})})- \frac{\p}{\p \tau}(\frac{\p L'}{\p
(\frac{\p q_{i}}{\p \tau})})=0.
\end{equation}
Calculations show that eqn. (28) leads to well-known Lagrangian
equation as
\begin{equation}
\frac{\p {\cal{L}}}{\p q_{i}} - \frac{d}{dt}(\frac{\p
{\cal{L}}}{\p (\frac{dq_{i}}{dt})})=0.
\end{equation}

Using eqn. (20), we have
\begin{equation}
H_{t}=- {\cal L} + \frac{\p \cal L}{\p \dot{q_{i}}}\dot{q_{i}},
\end{equation}
In order to have a consistent theory, one should consider the
total variation of $H_{t}$. In fact
\begin{equation}
dH_{t}=-\frac{\p \cal{L}}{\p t} dt.
\end{equation}
Making use of eq. (10), one finds
\begin{equation}
dH_{t}=- \frac{\p L'}{\p t}d\tau.
\end{equation}
Besides, the quantity $H_{0}$ is identically satisfied and does
not lead to constriants.

One should notice that equations (21,22) are equivalent to
equations (28,29).
\section{Classical fields as constrained systems}
In the following sections we would like to study the Hamiltonian
and the Lagrangian formulations for classical field systems and
demonstrating the equivalence between these two formulations for
the reparametrization invariant fields.

 A classical relativistic field $ \f_{i}= \f_{i}(\vec{x}, t)$ in four space-time dimensions may be
described by the action functional
\begin{equation}
S(\f_{i})= \int dt \int d^{3}x\{{\cal{L}}(\f_{i},
\p_{\m}\f_{i})\},\;\;\m= 0, 1, 2, 3,\;\;i = 1, 2,...,n,
\end{equation}
which leads to the Euler-Lagrange equations of motion as
\begin{equation}
\frac{\p {\cal{L}}}{\p {\f}_{i}} -\p_{\m}[\frac{\p
{\cal{L}}}{\p(\p_{\m}\f_{i})}]=0.
\end{equation}

One can go over from the Lagrangian description to the
Hamiltonian description by using the definition

\begin{equation}
\pi_{i}=\frac{\p \cal{L}}{\p{\dot{\f_{i}}}},
\end{equation}
then canonical Hamiltonian is defined as
\begin{equation}
H_{0}=\int d^{3}x(\pi_{i} {\dot{\f_{i}}} -\cal{L}).
\end{equation}
The equations of motion are obtained as \bn &&{\dot{\pi_{i}}} =
-\frac{\p H_{0}}{\p \f},\\
&&{\dot {\f}}= \frac{\p H_{0}}{\p {\pi}_{i}}. \en

\section{Reparametrization invariant fields}

In analogy with the finite dimensional systems, we introduce the
reparametrization invariant action for the field system as
\begin{equation}
S=\int d\tau\int {\cal{L}}_{R} d^{3}x,
\end{equation}
where
\begin{equation}
{\cal{L}}_{R}= {\dot{t}}{\cal{L}}(\f_{i}, \p_{\m}\f_{i}).
\end{equation}

Following the canonical method [1-4], we obtain the set of [HJPDE]
as \bn&& {H'}_{0}=\pi_{\tau}+ \pi_{i}^{(\tau)} \frac{d
{\f}_{i}}{d\tau} +\pi_{t}\frac{dt}{d\tau} -{\cal{L}}_{R}= 0,
\;\;\pi_{\tau}= \frac{\p S}{\p \tau},\\
&&{H'}_{t}= \pi_{t} + H_{t} =
0,\;\;\;\;\;\;\;\;\;\;\;\;\;\;\;\;\;\;\;\;\;\;\;\;\;\;\;\;\;\;\;
\pi_{t}=
\frac{\p S}{\p t}, \en where $H_{t}$ is defined as
\begin{equation}
H_{t}= -{\cal{L}}(\f_{i}, \p_{\m}\f_{i}) + \pi_{i}^{(\tau)}\frac{d
{\f}_{i}}{dt},
\end{equation}
and $\pi_{i}^{(\tau)}$, $\pi_{t}$ are the generalized momenta
conjugated to the generalized coordinates $\f_{i}$ and $t$
respectively.

The equations of motion are obtained as \bn&&d\f_{i}= \frac{\p
{H'}_{0}}{\p \pi_{i}}d\tau + \frac{\p {H'}_{t}}{\p \pi_{i}}dt=
\frac{\p {H'}_{t}}{\p \pi_{i}}dt,\\
&&d\pi^{i}=- \frac{\p {H'}_{0}}{\p \f_{i}}d\tau - \frac{\p
{H'}_{t}}{\p \f_{i}}dt= -
\frac{\p {H'}_{t}}{\p \f_{i}}dt,\\
&&d\pi_{t}=- \frac{\p {H'}_{0}}{\p t}d\tau - \frac{\p
{H'}_{t}}{\p t}dt= 0. \en

Now the Euler-Lagrangian equation for the field system reads as
\begin{equation}
\frac{\p {\cal{L}}}{\p \f_{i}} - \frac{\p}{\p x^{\m}}(\frac{\p
{\cal{L}}}{\p (\frac{\p \f_{i}}{\p x_{\m}})})=0.
\end{equation}

Again as for the finite dimensional systems, equations (44,45) are
equivalent to equations (47) for field systems.
\section{ Conclusion}
As it was mentioned in the introduction, if the rank of the
Hessian matrix for discrete systems is $(n-r)$; $0< r< n$, then
the systems can be treated as field systems [5]. The treatment of
Lagrangians as field systems is always in exact agreement with
the Hamilton-Jacobi treatment for reparametrization invariant
theories. The equations of motion (21, 22) are equivalent to the
equations of motion (28, 29). Besides the the variations of
constraints (31) and (32) are identically satisfied and no
further constraints arise.

In analogy with the finite dimensional systems, it is observed
that the Lagrangian and the Hamilton-Jacobi treatments for the
reparametrization invariant fields are in exact agreement.

\end{document}